# Three Approaches to Probability Model Selection

**William B. Poland**
Strategic Decisions Group
2440 Sand Hill Road
Menlo Park, CA 94025-6900
*poland@leland.stanford.edu*

**Ross D. Shachter**
Department of Engineering-Economic Systems
Stanford University
Stanford, CA 94305-4025
*shachter@camis.stanford.edu*

## Abstract

This paper compares three approaches to the problem of selecting among probability models to fit data: (1) use of statistical criteria such as Akaike's information criterion and Schwarz's "Bayesian information criterion," (2) maximization of the posterior probability of the model, and (3) maximization of an "effectiveness ratio" trading off accuracy and computational cost. The unifying characteristic of the approaches is that all can be viewed as maximizing a penalized likelihood function. The second approach with suitable prior distributions has been shown to reduce to the first. This paper shows that the third approach reduces to the second for a particular form of the effectiveness ratio, and illustrates all three approaches with the problem of selecting the number of components in a mixture of Gaussian distributions. Unlike the first two approaches, the third can be used even when the candidate models are chosen for computational efficiency, without regard to physical interpretation, so that the likelihoods and the prior distribution over models cannot be interpreted literally. As the most general and computationally oriented of the approaches, it is especially useful for artificial intelligence applications.

## 1 INTRODUCTION

The model selection problem is to select from a candidate set of models the one that fits the input data the "best" in some sense. There is a large body of literature on model selection; see for example Fourth International Workshop on Artificial Intelligence and Statistics [1993], for which model selection was the primary theme, and Ljung [1987]. This paper considers only approaches that can be interpreted as maximizing a penalized likelihood function, ignoring other important approaches such as Laplace approximation of the posterior probabilities of models [Kass and Raftery 1993] and minimum description length criteria [Rissanen 1985]. Also, this paper emphasizes the case that the candidate models form a sequence of nested (successively more general) parametric, continuous probability distributions with increasing numbers of parameters. An example of such a sequence, used throughout, is the sequence of mixtures of Gaussian distributions (Gaussian mixtures) with increasing numbers of components, starting with one component. A mixture distribution with m continuous components has a density of the form

$$f(x) = p_1 f_1(x) + \cdots + p_m f_m(x) \qquad (1)$$

where $p_1, \cdots, p_m$ are positive numbers summing to one and $f_1(x), \cdots, f_m(x)$ are the component densities. Mixtures of analytically tractable component distributions, such as Gaussians, are useful to model not only true mixtures but any continuous probability distributions with which fast calculations are desirable [Poland 1994].

Section 2 reviews a class of large-sample statistical criteria for model selection, sometimes called generalized information criteria. Based on limiting properties of the maximum likelihood as the sample size increases, these criteria specify maximization of the log-likelihood of the sample, less a penalty increasing in the number of independent model parameters. The log-likelihood for each model is evaluated at the maximum-likelihood estimates of the model's parameters. The criteria include the well-known Akaike's information criterion (AIC) and Schwarz's "Bayesian information criterion" (BIC). Stone [1977] showed that the AIC is asymptotically equivalent under certain conditions to model choice by cross-validation, which evaluates the likelihood of each sample point based on all other points rather than all points. The BIC assumes unspecified priors for the parameters meeting some weak conditions, and is similar in effect to the minimum description length principle, which requires minimizing the number of bits of information needed to represent the data [Sclove 1993].

Unfortunately, mixture distributions with a finite number of components do not meet the regularity conditions assumed in the expansions used to derive these criteria [Sclove 1993, Titterington et al. 1985]. However, this problem can be avoided in practice, and Sclove reported that the criteria have met with good success in estimating the number of components of a mixture. This is to be expected given their equivalence, under some conditions, to the approaches of Sections 3 and 4.

Section 3 presents a maximum a posteriori (MAP) approach to model selection: to specify a prior distribution over the candidate models and then select the most probable model posterior to the data. This is a practical simplification of a full-fledged Bayesian approach, which would keep all the candidate models weighted by their posterior probabilities. Section 3 also re-derives the equivalence, noted by Sclove [1993], of the MAP approach with suitable prior distributions to the statistical criteria approach.

Section 4 presents a third approach: to maximize the ratio of the likelihood of the data, as an accuracy measure, to a computational cost measure that increases with the number of model parameters [Poland and Shachter 1993]. For a certain form of the cost function, this approach is shown to be equivalent to the MAP approach. However, for another, reasonable form of the cost function, it always applies a smaller penalty for sufficiently many parameters.

This effectiveness ratio has the advantage of remaining easy to interpret in the common situation that the candidate models are not believed to be literally correct or representative, so that the likelihoods, and the prior for the MAP approach, cannot be interpreted literally. For example, the candidate models might be given a simple parametric form only to allow efficient computations, as long as the computations do not depend on this internal form. A second, related advantage of the approach is that it allows a continuous distribution rather than a sample as the input data: that is, given the correct continuous distribution, the model selection process can find a more computationally convenient representation. In this case, the accuracy measure must be generalized from a likelihood to a function of the *relative entropy* from the input distribution to the modeled distribution. Relative entropy, also known as Kullback-Leibler number, directed divergence, cross-entropy, etc., is a directed measure of the divergence between two continuous or two discrete probability distributions; see Shore [1986] and Whittaker [1990]. Poland and Shachter [1993] used this approach to fit Gaussian mixture distributions to arbitrary continuous input distributions. The EM (expectation-maximization) algorithm determined the parameters that minimize the relative entropy for any given number of components; then maximization of the effectiveness ratio determined the appropriate number of components.

Section 5 presents a heuristic to simplify the search for the best model, and compares the results of all three approaches in a mixture selection example. Section 6 concludes that the effectiveness ratio approach can be especially useful in artificial intelligence applications, because it accounts for computational cost explicitly and does not require belief in the validity of one of the candidate models.

## 2 SOME STATISTICAL CRITERIA FOR MODEL SELECTION

A class of statistical criteria for model selection, given a large sample $x = (x_1, \cdots, x_n)$, has the form

$$\hat{m} = \arg\max_{m = 1,2,\ldots} \left[ \ln f^{ML}_{(m)}(x) - c(n)\, v(m) \right] \quad (2)$$

where m indexes the candidate models, $f^{ML}_{(m)}(x)$ is the likelihood for model m with maximum-likelihood parameters, c(n) is a quantity depending only on n and the criterion, and $v(m)$ is the number of independent parameters in model m. For example, $v(m) = 3m - 1$ for a Gaussian mixture with m univariate components, because each component has an associated weight, mean, and variance, and the weights (which are probabilities) are constrained to sum to one. The AIC sets c(n) = 1, whereas the BIC sets $c(n) = 1/2 \ln n$. Therefore the BIC favors smaller models than the AIC for $1/2 \ln n > 1$, or $n \geq 8$.

For successive nested models, the likelihood function must be nondecreasing, because it is a maximum over successively more general models. Therefore the second term of (2) is needed to limit the size of the selected model, and it can be viewed as a penalty for the model size.

## 3 A MAP APPROACH

A full-fledged Bayesian approach to the problem of model selection is not to select any single model, but to specify a prior distribution over a mutually exclusive and collectively exhaustive set of models, and to update this with the data. (Thus the prior and posterior distributions for the uncertain model would be mixture distributions, but should not be confused with the models when they are themselves mixture distributions.) All models with nonzero posterior probability would be used in subsequent analysis. This is of course impractical, so a common alternative is to select a *single* model with maximum a posteriori (MAP) probability. (See Draper [1993] for a case for compromising between retaining all models and selecting the single most probable model.)

Typically, each model has some uncertain parameters. A statistical approach to this intra-model uncertainty begins by expressing the probability of the data given a model as an integral, over the model's parameters, of this probability *given* the parameters times the probability of the parameters given the model. Then large-sample Laplace approximations of this integral lead to criteria such as the BIC [Draper 1993]. A simpler approach to intra-model uncertainty is taken here, which does not require a large sample: just as we select the MAP model, we select the MAP parameters for each model. For example, MAP estimates of the parameters of a Gaussian mixture distribution with a given number of components can be found by the EM algorithm for maximum-likelihood estimation, augmented with simple priors for the component parameters [Hamilton 1991]. (Note that MAP estimation of the parameters reduces to maximum-likelihood estimation when the joint prior distribution for the parameters is uniform over the region of interest.)

When the candidate models are nested, as for mixtures of one or more Gaussians, they will not be mutually exclusive, as assumed for the specification of the prior distribution over models. However, the models will be mutually exclusive after MAP selection of their parameters, except in the rare case that the MAP parameters of two models match. In this case one of the models can be dropped and its probability added into that of the other model.

A convenient choice for the prior distribution over the candidate models, when they can be indexed by the counting numbers, is the geometric distribution, the discrete analog of the exponential distribution:

$$P(M = m) = p_1 (1 - p_1)^{m-1}, \quad m = 1, 2, \cdots \quad (3)$$

where $p_1$ is the probability that m is one. This makes successively larger models less probable by a constant factor. As $p_1$ approaches zero, the geometric distribution approaches an improper uniform distribution over the counting numbers, which conveniently makes MAP estimation reduce to maximum-likelihood estimation. Thus, the parameter of the geometric distribution gives full control over the rate that the probabilities of successive models decrease.

The MAP model maximizes the product of the prior and the likelihood. Letting $f_{(m)}^{MAP}(x)$ be the likelihood for model m with MAP parameters, this maximization with a geometric prior over models is

$$\hat{m} = \underset{m = 1,2,\ldots}{\arg\max} \; [\, p_1 (1 - p_1)^{m-1} f_{(m)}^{MAP}(x) \,]$$
$$= \underset{m = 1,2,\ldots}{\arg\max} \; \left[\, \ln f_{(m)}^{MAP}(x) - \left(\ln \frac{1}{1-p_1}\right)m \,\right]. \quad (4)$$

As in (2), the second term can be viewed as a penalty for the model size. In fact, comparison of (4) and (2) shows that the MAP approach with the geometric prior on models becomes equivalent to the statistical criteria approach when (i) the joint prior for the parameters of each model is uniform over the region of interest, so that $f_{(m)}^{MAP}(x)$ is the same as $f_{(m)}^{ML}(x)$, and (ii) $\{\ln[1/(1-p_1)]\}m$ is equal to $c(n)v(m)$, except for terms independent of m that would not affect the maximization.

Letting $v'(m)$ represent $v(m)$ less any terms independent of m, this second requirement gives

$$p_1 = 1 - e^{-c(n) v'(m) / m}. \quad (5)$$

Gaussian mixtures have $v'(m) = 3m$. Therefore, for Gaussian mixtures the AIC, for which $c(n) = 1$, effectively sets $p_1 = 1 - e^{-3} \approx 0.95$, while the BIC, for which $c(n) = 1/2 \ln n$, effectively sets $p_1 = 1 - n^{-3/2}$, which is 0.999 for $n = 100$, for example. Thus the AIC and BIC applied to nested probability models have the same effect as MAP estimation with very high probabilities for the smallest models, despite the differences in the origins of the approaches. If the prior distribution over models were more complicated than the geometric—for example, with a mode after the smallest model—we might still expect geometric tail behavior and thus obtain similar correspondences for large enough models.

## 4 AN EFFECTIVENESS RATIO APPROACH

A third approach to probability model selection is to maximize an "effectiveness ratio": the likelihood for each model, as an accuracy measure, divided by a computational cost measure. (Similar objectives are useful in search problems; see Simon and Kadane [1975] for an example.) The parameters of each model can be maximum-likelihood or other estimates, as long as the resulting accuracy measure is useful. Letting $\hat{f}_{(m)}(x)$ represent the estimated likelihood and g(m) represent the cost measure for model m, this approach seeks

$$\hat{m} = \underset{m = 1,2,\ldots}{\arg\max} \; [\, \hat{f}_{(m)}(x) / g(m) \,]$$
$$= \underset{m = 1,2,\ldots}{\arg\max} \; [\, \ln \hat{f}_{(m)}(x) - \ln g(m) \,]. \quad (6)$$

For nested models the cost measure can be viewed as a size penalty. One possibility is $g(m) = a \, k^m$, where a is a positive constant that need not be assessed and k is a parameter greater than one. Comparing (6) with (4) shows that maximizing the effectiveness ratio with this cost measure, and with MAP estimates of the model parameters, is equivalent to the MAP approach with a geometric($p_1$) prior such that $p_1 = 1 - 1/k$. Conversely, the MAP approach with a geometric($p_1$) prior can be viewed as maximizing an effectiveness ratio with cost proportional to $[1/(1 - p_1)]^m$, and with MAP estimates of the model parameters.

Another reasonable cost measure, used in the example in the next section, is $g(m) = a \, m^k$, where again a is a positive constant that need not be assessed, and k is a positive parameter. A possible setting for k in a probability problem with only discrete (including discretized) and mixture variables is the total number of variables in the problem; then $m^k$ is the worst-case number of combinations of possible discrete outcomes and mixture components needed to analyze the problem, if all variables have the same number, m, of possible outcomes or components. With this cost function, (6) becomes

$$\hat{m} = \underset{m = 1,2,\ldots}{\arg\max} \; [\, \ln \hat{f}_{(m)}(x) - k \ln m \,]. \quad (7)$$

In two important cases, effectiveness ratio maximization remains straightforward to interpret when the other approaches do not. First, we may not believe *any* of the candidate models, having chosen them only for analytical convenience. Then the likelihoods and prior used in the first two approaches have no literal meaning, but the likelihoods can still serve as accuracy measures in the effectiveness ratios. Second, we might want to select and *fit* an approximate model to a given continuous input distribution, rather than select and *estimate* a model from a given sample. Then a natural choice of accuracy measure is a generalization of the log-likelihood for an exchangeable sample, in terms of the relative entropy from the input distribution to the model distribution. The rest of this section derives this generalization [Kullback 1968, Titterington et al. 1985] and applies it to (6).

The *relative entropy* from a continuous random variable X to another continuous random variable Y can be defined as an expectation over X:

$$D(X,Y) = E\{ \ln[ f_X(X) / f_Y(X) ] \}$$
$$= \int_{-\infty}^{\infty} \ln[ f_X(x) / f_Y(x) ] f_X(x) \, dx \quad (8)$$

where $f_X(\cdot)$ and $f_Y(\cdot)$ are the densities of X and Y respectively. (The expectation operator applies to the argument of both density functions.) Relative entropy is nonnegative and is zero only when the two densities are equal everywhere. The *entropy* of a continuous random variable X is defined as

$$H(X) = -E[ \ln f_X(X) ], \quad (9)$$

so relative entropy can be expressed as a difference:

$$D(X,Y) = -E[ \ln f_Y(X) ] - H(X). \quad (10)$$

Let X represent a random draw from the exchangeable sample $x = (x_1, \cdots, x_n)$: that is, X equals $x_i$ with probability $1/n$ for $i = 1, \cdots, n$. Let Y have the probability distribution that generates each element of the sample, and let **Y** have the corresponding joint distribution. Then in general the likelihood of **x**, $f_\mathbf{Y}(\mathbf{x})$, can be expressed in terms of an expectation over X:

$$f_\mathbf{Y}(\mathbf{x}) = \prod_{i=1}^{n} f_Y(x_i)$$
$$= \exp[ \ln \prod_{i=1}^{n} f_Y(x_i) ]$$
$$= \exp[ \sum_{i=1}^{n} \ln f_Y(x_i) ]$$
$$= \exp[ n \sum_{i=1}^{n} \frac{1}{n} \ln f_Y(x_i) ]$$
$$= \exp\{ n E[ \ln f_Y(X) ] \}. \quad (11)$$

Therefore the effectiveness ratio maximization (6) can be expressed in terms of X as

$$\hat{m} = \arg\max_{m=1,2,\ldots} \{ n E[ \ln \hat{f}_{(m)}(X) ] - \ln g(m) \}$$
$$= \arg\max_{m=1,2,\ldots} \{ E[ \ln \hat{f}_{(m)}(X) ] - \frac{1}{n} \ln g(m) \}. \quad (12)$$

If n is large, we might want to approximate the distribution of X with a simpler one that groups the $x_i$ into categories and assigns them differing probabilities. We might even approximate X as continuous; the expectation notation of (12) is general enough to accommodate this. Moreover, if X is in fact continuous—that is, if we seek the best fit to a given continuous distribution—we can still use (12) by viewing the continuous distribution as equivalent to an unspecified exchangeable sample of some specified size, n. For example, if we elicit a judgmental cumulative probability distribution for X from an expert, we might set n equal to the number of cumulative points elicited. The parameter n reflects our uncertainty about the input distribution; if there is none, n is infinite and (12) requires a perfect fit. Conveniently, with the cost measure $g(m) = a \, m^k$ as in (7), (12) requires only the combined parameter $k/n$.

With X continuous, the expected log-likelihood in (12) can be found from (10), with the entropy and relative entropy calculated numerically. However, since the entropy H(X) is independent of m, (12) can be expressed directly in terms of relative entropy:

$$\hat{m} = \arg\min_{m=1,2,\ldots} \left[ D(X,\hat{Y}_{(m)}) + \frac{1}{n} \ln g(m) \right] \quad (13)$$

where $\hat{Y}_{(m)}$ has the density estimated for model m. The example below illustrates this minimization of relative entropy plus the size penalty.

## 5   A SEARCH HEURISTIC AND A COMPARISON USING MIXTURE MODELS

A simple heuristic to search for the best model from a sequence of nested models, for any of the three approaches, is to try successive models, starting with the smallest, and stop with model m when the objective function for model m + 1 is no greater than for model m. This heuristic could miss the global optimum if the objective function has multiple local optima, so we might check one or more subsequent models to reduce this risk. (Cheeseman et al. [1988] give a similar heuristic that decrements the model index.) Such a check is unnecessary in the example below.

For effectiveness ratio maximization, this heuristic stops when

$$E[ \ln \hat{f}_{(m+1)}(X) ] - \frac{1}{n} \ln g(m+1)$$
$$\leq E[ \ln \hat{f}_{(m)}(X) ] - \frac{1}{n} \ln g(m)$$

or

$$E[ \ln \hat{f}_{(m+1)}(X) ] - E[ \ln \hat{f}_{(m)}(X) ]$$
$$\leq \frac{1}{n} \ln \frac{g(m+1)}{g(m)}. \quad (14)$$

If X is continuous, this heuristic can simply test the decrease in relative entropy when the model index is incremented:

$$D(X,\hat{Y}_{(m)}) - D(X,\hat{Y}_{(m+1)}) \leq \frac{1}{n} \ln \frac{g(m+1)}{g(m)} \quad (15)$$

where $\hat{Y}_{(m)}$ has the density estimated for model m. This decrease in relative entropy is nonnegative because the models are nested, allowing successively better fits. With the size penalty $g(m) = a k^m$, the right-hand side of (15) becomes the constant $(\ln k) / n$, while with $g(m) = a m^k$, as in the example below, the right-hand side of (15) becomes $k/n \ln[(m + 1)/m]$, a slowly increasing function of m.

Figure 1 illustrates all three approaches. The candidate models are mixtures of one or more univariate Gaussian distributions with parameters determined by the EM algorithm, which part a illustrates. The input data is an exponential distribution (chosen for its simplicity and very non-Gaussian nature), viewed as equivalent to an exchangeable sample of size n = 100. Since we already know the continuous distribution, the model selection problem is to select the Gaussian mixture that best fits this distribution, accounting for the anticipated cost of calculations using the result.

The continuous input allows all three methods to be viewed as minimizing relative entropy plus a size penalty. This penalty is
- $3/n$ m for the AIC,
- $3/2 [(\ln n)/n]$ m for the BIC,
- $\{ \ln[1/(1-p_1)] / n \}$ m for the MAP approach with a geometric($p_1$) prior, and
- $k/n \ln m$ for effectiveness ratio maximization with cost proportional to $m^k$.

Note that the penalty always decreases (that is, the accuracy term becomes relatively more important) with the sample size n, but more slowly for the BIC than for the other objectives.

Part b indicates the similarity of the approaches for some parameter settings. However, as part b suggests, for large sizes the size penalty for effectiveness ratio maximization becomes weaker than for the other approaches (except the MAP approach with a flat prior, which requires a perfect fit), because the penalty is logarithmic rather than linear in m. The optimal number of components is infinite when there is no penalty, but is only two to six for the other objectives illustrated. A uniform, rather than exponential, input distribution gives similar results with one to six components optimal [Poland 1994], and we can expect similarly parsimonious representations with arbitrary input distributions, allowing simplified calculations.

## 6 CONCLUDING REMARKS

As Figure 1 shows, all three approaches can behave similarly. However, effectiveness ratio maximization has several advantages. It is the most general approach. It accepts a continuous distribution as its input, as well as a sample from one. It is meaningful when none of the candidate models is believed to be correct or representative of reality, as when the modeling process is driven by computational cost considerations. Finally, it accounts for computational cost explicitly. These characteristics are especially valuable in artificial intelligence applications in which accuracy and computational cost need to be traded off automatically in selection from a limited set of candidate probability models.

Computational experience with full-scale applications would provide valuable feedback about the relative merits of the approaches and the most useful extensions. These might include more general model selection objectives than the effectiveness ratio and more flexible prior distributions over the model size than the geometric. Cases with little data, violating the large-sample assumption of the statistical criteria, would be of particular interest.

### Acknowledgments

This research benefited from discussions with our colleagues in the Engineering-Economic Systems Department and Geoffrey Rutledge of the Section on Medical Informatics, as well as suggestions from the referees.

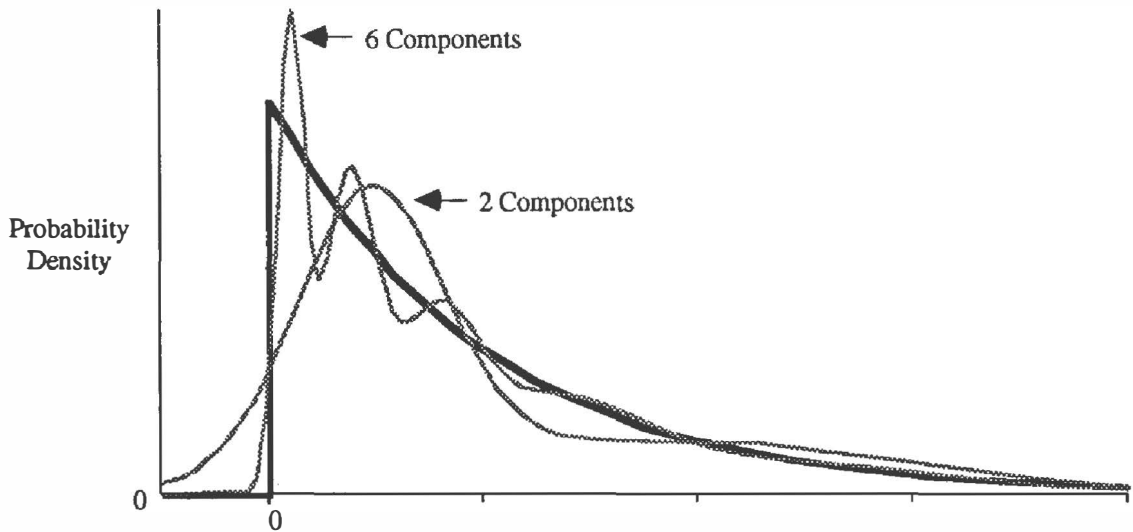

(a) Two Gaussian Mixtures Fitted to an Exponential Distribution

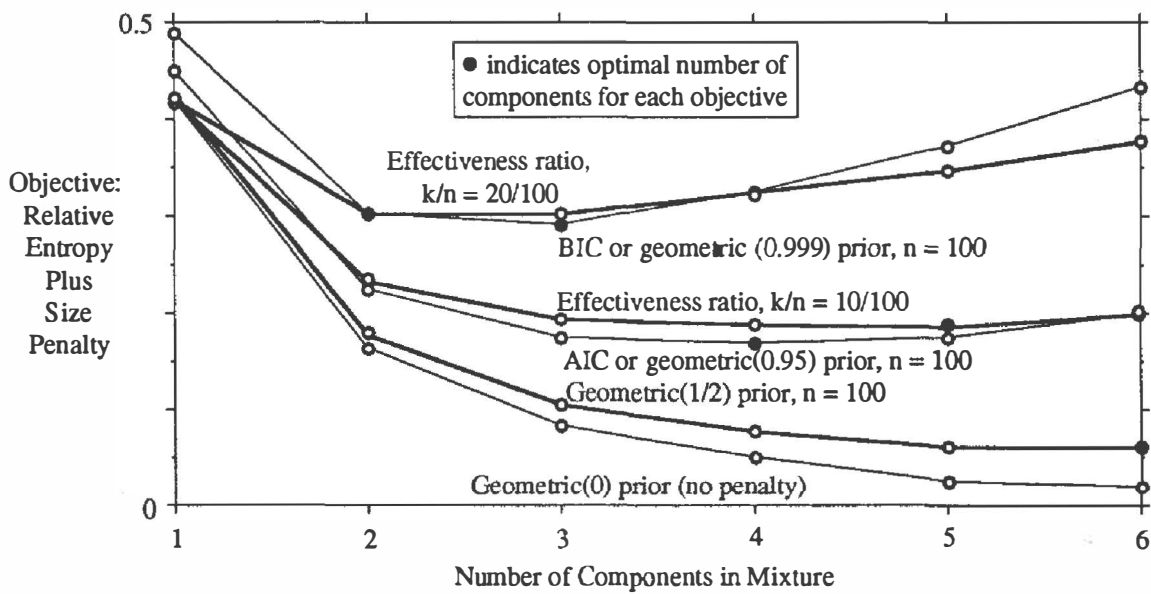

(b) Alternative Objective Functions to be Minimized

Figure 1: Selecting a Gaussian Mixture Model for an Exponential Distribution